\documentclass[apjl,twocolumn]{emulateapj}

\usepackage[pagebackref,breaklinks,colorlinks,citecolor=blue]{hyperref}

\usepackage{amsmath}
\usepackage{amssymb}

\usepackage{graphicx}
\usepackage{color}
\usepackage{natbib}
\usepackage{xspace}
\usepackage{nicefrac}
\usepackage{gensymb}
\usepackage[dvipsnames]{xcolor}
\newcommand\myshade{85}
\usepackage[T1]{fontenc}
\usepackage{lmodern}

\definecolor{aqua}{rgb}{0.2, 0.7, 0.7}
\definecolor{brightgreen}{rgb}{0.4, 1.0, 0.0}
\definecolor{darkpastelgreen}{rgb}{0.01, 0.75, 0.24}
\definecolor{tealblue}{rgb}{0.21, 0.46, 0.53} 	
\definecolor{pansypurple}{rgb}{0.47, 0.09, 0.29}
\definecolor{cinnabar}{rgb}{0.89, 0.26, 0.2}
\definecolor{myblue}{RGB}{35,107,142}

\colorlet{mylinkcolor}{aqua}
\colorlet{mycitecolor}{aqua}

\hypersetup{
  citecolor  = mycitecolor!\myshade!black,
  linkcolor  = mylinkcolor!\myshade!black,
  colorlinks = true,
}

\newcommand{\Msun}{M_\odot \xspace}
\newcommand{\Rsun}{R_\odot \xspace}

\def\Msun{{\rm M}_{\odot}} 
\def\Rsun{{\rm R}_{\odot}}

\def\aj{AJ}                 
\def\apj{ApJ}                
\def\apjl{ApJL}             

\def\mnras{MNRAS}            
\def\prd{Phys.~Rev.~D}       
\def\nat{Nature}              
\def\aap{A\&A}                

\def\nar{New Astronomy Reviews}

\submitted{Accepted for publication in Astrophysical Journal Letters: March 18, 2017}

\begin{document}

\title{Electromagnetic signals following stellar--mass black hole mergers }
\shorttitle{Electromagnetic signals following stellar--mass black hole mergers }

\author{S. E. de Mink\altaffilmark{1,2} \&  A. King\altaffilmark{3,1,4}} 

 \affil{ $^{1}$Anton Pannekoek Institute for Astronomy, University of
Amsterdam, 1090 GE Amsterdam, The Netherlands \\
$^{2}$Kavli Institute for Theoretical Physics, University of California, Santa
Barbara, CA 93106, USA \\
$^{3}$ Theoretical Astrophysics Group, Department of Physics \& Astronomy, University of Leicester, Leicester LE1 7RH, UK\\
$^4$ Leiden Observatory, Leiden University, Niels Bohrweg 2, NL-2333 CA Leiden, The Netherlands\\  
email: S.E.deMink@uva.nl,  ark@leicester.ac.uk  
} 

\begin{abstract}
It is often assumed that gravitational wave~(GW) events resulting from the merger of stellar--mass black holes are unlikely to produce electromagnetic~(EM) counterparts.  We point out that the progenitor binary has probably shed  a mass $\gtrsim 10\,{\rm M}_{\odot}$
during its prior evolution. If a tiny fraction of this gas is retained until the merger, the recoil and sudden mass loss of the merged black hole shocks and heats it within hours of the GW event. Whether the resulting EM emission is detectable uncertain. The optical depth through the disk is likely to be high enough that the prompt emission consists only of photons from its optically thin skin, while the majority may take years to emerge. However, if some mechanism can release more photons in a time comparable to the few--hour energy production time, the peak luminosity of the EM signal could be detectable. For a disk retaining only $\sim 10^{-3}$ of the mass shed in the earlier binary evolution,
medium--energy X--rays to infrared emission would be observable hours after the GW event
for source
distances  $\sim 500\,\rm{Mpc}$. 
Events like this may already have been observed, but ascribed to unidentified active galactic nuclei. Improved sky--localization should eventually allow identification based on spatial coincidence. A detection would provide unique constraints on formation scenarios and
potentially offer tests of strong--field general relativity.
 Accordingly we argue that the high scientific payoff of an EM detection fully justifies search campaigns. 
\end{abstract}


\section{Introduction}

The direct detection of gravitational waves from binary black hole mergers GW150914, GW151224, and a possible third event LVT151012, has drawn wide attention \citep{LIGO+2016_mainPaper, Ligo+2016_BoxingDayEventGW151226, Ligo_2016_BBHinfirstRun}.  The general anticipation that the first events would involve neutron stars rather than black holes mobilized massive coordinated campaigns to search for accompanying electromagnetic (EM) counterparts  \citep[e.g.][]{LIGO+2016_EMFollowup_GW150914}. For merger events involving at least one neutron star, a plethora of accompanying EM signatures is expected at a range of timescales and wavelengths. In contrast, for binary black hole mergers the common consensus has been that no significant EM counterpart is expected, except for "those in highly improbable environments pervaded by large ambient magnetic fields or baryon densities" as \citet{LIGO+2016_EMFollowup_GW150914} state.

Understandably, the report of a transient signal detected by the Fermi Gamma-ray Burst Monitor 0.4 seconds after the first event \citep{Connaughton+2016} attracted considerable attention. It encouraged several theoretical speculations for a possible origin of this unanticipated EM signal. But the lack of a corresponding detection by INTEGRAL/SPI-ACS \citep{Savchenko+2016}, careful reanalysis of the data \citep{Xiong2016} and reassessment of the low count statistics \citep{Greiner+2016} all lead to the conclusion that the Fermi trigger is consistent with a background fluctuation, and unlikely to be of astrophysical origin.  
This is perhaps not surprising: the ultra--prompt nature of the Fermi signal implies an extremely small EM source and probably the near--simultaneous formation of the second black hole, ruling out the usual formation channels. Proposed scenarios include the exotic formation of a binary black hole inside a massive star \citep{Loeb2016}, which has been ruled implausible \citep{Woosley2016,Dai+2016}. Alternatively, the survival of a minidisk surrounding one of the individual black holes activated shortly before the merger \citep{Perna+2016}  implies timescales and luminosities inconsistent with the Fermi signal \citep{Kimura+2016}.  Nevertheless, the question whether binary black hole mergers can have EM counterparts is natural and interesting.

Here we question the common consensus that the typical stellar--mass binary black hole merger is always dark. We consider a simple possibility for EM signals following binary black hole mergers, with a possible delay of hours. This mechanism requires the BH binary to have a
circumbinary disk at the time of merger whose mass need only be a very small fraction of that shed as the system evolved. In \autoref{sec:form} we argue that the initial formation of such a disk is natural in the two progenitor formation scenarios for isolated binary evolution, but unlikely if the binary formed through dynamical interactions in a cluster. We discuss the evolution of the disk under viscous and tidal stresses. If any such disk survives until coalescence, it is perturbed both by the GW recoil and by the sudden drop in mass as energy and momentum is radiated away by gravitational waves.  In  \autoref{sec:pred} we use simple analytic scaling arguments, calibrated against earlier detailed simulations for supermassive black hole mergers, to provide crude estimates of the possible luminosity, delay time, duration and approximate spectral energy distribution of the resulting signal.  In \autoref{sec:dis} we emphasize that the detectability of these signals is subject to large uncertainties; especially the survival of the disk until the merger, and how promptly photons  escape it afterwards.  If the EM signals are detectable, coordinated searches should allow us to distinguish them from background contamination by active galactic nuclei, given the anticipated improvements of sky localization and possibly advance detection by eLISA several years before coalescence. Detection or 
non--detection of EM signals will contribute unique insight into the earlier evolution of the merging black hole binaries, testing for circumbinary material and any post--merger recoil.  The latter would allow new direct tests of strong--field general relativity.

\begin{figure}\center
  \includegraphics[width=0.4\textwidth]{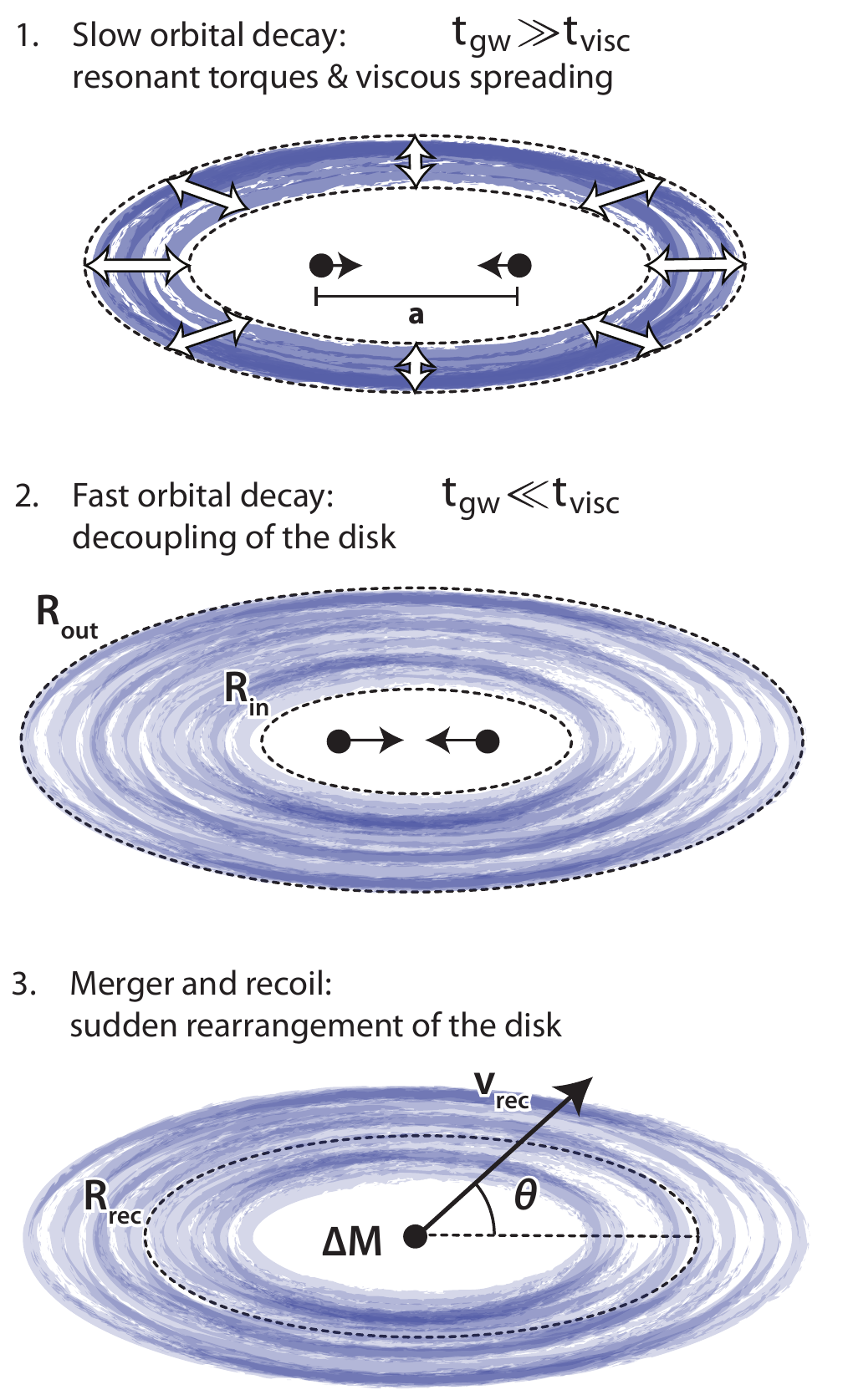}
  \caption{ Cartoon depicting the evolution of a circumbinary disk.  \label{scenario}}
\end{figure}

\section {Circumbinary disk}  \label{sec:form}

\paragraph{Origin \label{sec:form1}}

Circumbinary material is found around stellar binary systems in a wide variety of evolutionary phases. Examples include very young systems where the disk may be a remainder of the formation process \citep[e.g.][]{Rosenfeld+2013, Janson+2016}.  In more evolved systems, circumbinary disks can be formed by mass loss from one or both stars through the outer Lagrangian points, as seen in recent simulations \citep[e.g.][]{Pejcha+2016, Kuruwita+2016}.  Observed examples include the subdwarf + Be star system HR 2142 \citep{Peters+2016}, the massive system RY Scuti \citep{Smith+2002} and various post asymptotic giant branch stars including the famous Red Rectangle  \citep[e.g.][]{Cohen+2004, Hillen+2016}.   The presence of circumbinary material has also been considered for the microquasar SS 433, which may contain a black hole of about $10\Msun$ \citep{Perez-M.+2010, Bowler2013}.

The progenitors of binary black holes shed mass during various stages of their evolution. Fast ejecta, such as mass loss by radiatively driven stellar winds \citep[e.g.][]{Vink+2000} are likely to escape the system.  But most mass leaves the system with velocities comparable to or below the orbital velocities, and so stays close to the binary at least initially. The most obvious example is common--envelope ejection \citep{Ivanova+2013}.  Other examples include eruptive mass--loss episodes \citep{Humphreys+1994}, material centrifugally shed from rapidly rotating stars after spin--up \citep{Krticka+2011}, mass shedding during non--conservative Roche--lobe overflow \citep{de-Mink+2009} and mass loss through a weak fallback supernova \citep[e.g.][]{Fryer2006, Lovegrove+2013}.  The amount of mass shed and the ejecta velocities are uncertain in each case, but generally expected to amount to many $\Msun$.
For example in the standard evolutionary path presented  by \citet{Belczynski+2016} for the first LIGO detection, about $90\Msun$ is lost from the system, about $45\Msun$ of it during the common envelope ejection.  Most promising  for our picture may be mass shed during the formation of the second black hole in a failed explosion. Mass that remains bound to the system is not exposed to the ionizing radiation and stellar winds of the progenitor stars.

We will argue in the next section that a long--lived disk of less than a percent of a solar mass already gives potentially detectable EM signals. The expected mass ejection in the scenario by \citet{Belczynski+2016}  is 3--4 orders of magnitude larger, so even a very inefficient mechanism for gas retention could in principle lead to observable signals. 
Gas retention seems most likely in formation channels for isolated binaries: (i) the \emph{classical common envelope scenario}, which involves multiple phases of highly non-conservative mass transfer  \citep[e.g.,][]{Tutukov+1973, Tutukov+1993, Kalogera+2007, Kinugawa+2016, Belczynski+2016} and the   (ii) \emph{chemically homogeneous evolution scenario}  in which two tidally distorted binary stars stay compact as they experience strong internal mixing \citep{de-Mink+2009a}, which has recently been proposed as a new channel for compact binary black hole formation \citep{Mandel+2016, Marchant+2016, de-Mink+2016}.  Circumbinary gas may also be expected for binary black hole pairs formed through the (iii) \emph{assisted inspiral in the disk of active galactic nuclei} \citep{Stone+2016,Bartos+2016}. 
In contrast, we do not expect circumbinary material to survive in the case of (iv) the \emph{dynamical formation channels}, where tight binary black holes are formed through multiple three body interactions in dense star clusters  \citep[e.g.][and subsequent papers]{Sigurdsson+1993, Kulkarni+1993}. The sequence of near collisions needed to produce a tight binary black hole will probably also strip it of any circumbinary material.   We conclude that EM signals from a surviving circumbinary disk can potentially distinguish between formation scenarios. 

\paragraph{Disk evolution}
The evolution and fate of a possible circumbinary disk surrounding two black holes is subject to substantial uncertainties and poses a hard computational problem.  Whether a disk can survive until coalescence of the black holes is a question that will ultimately be answered by observations. Here we assume that the disk has been formed by the progenitor stars through one of the scenarios above, and give simple physical considerations  specifying the evolution and possible long-term survival of a circumbinary disk. 

After the formation of the second black hole, there are no internal sources of stellar radiation. A fraction of the circumbinary disk may survive an initial phase of settling and cool to very low temperatures, probably even cooler than a normal protoplanetary disc \citep[$\sim$10 K, e.g.][]{Dullemond+2007}, implying an extremely small disc aspect ratio H/R $\sim 10^{-3}$. The disk mass is limited by self-gravity to a value H/R times the total binary mass $M$, otherwise fragmentation occurs \citep{Armitage+2001}.

Cold disks require an external ionizing source to activate the magneto--rotational instability and so evolve viscously.  A plausible source is the ambient cosmic ray flux. This ionizes a skin depth of order $\sim 100\, {\rm g\, cm^{-2}}$, giving a \citet{Shakura+1973} viscosity parameter $\alpha \sim 10^{-2}$ \citep{Gammie1996,Armitage+2001}  in the skin, where matter can at least initially flow inwards. The inflow rate is much lower near the mid plane of the disk, so there is a tendency for matter to accumulate in dead zones characterized by the  local grain opacities \citep{Gammie1996}.  

Adapting this picture to the circumbinary disk of a black hole binary is clearly complex. The initial conditions of the disk are very uncertain, and inflowing material is dammed up at orbital resonances with the central binary.  The inner region of the disk is subjected to resonant torques \citep[cf.][]{Lynden-Bell+1974, Pringle1991},  which opens an inner cavity with a radius of about $ R_{\rm in} \simeq 2a$, where $a$ is the binary separation \citep[e.g.]{Artymowicz+1994}.  

A crucial question is whether material can flow through the inner cavity. This is important since the resulting accretion luminosity can further ionize the disk. This would in turn activate the magneto--rotational instability and might potentially lead to a runaway effect emptying the disk through the inner cavity.  Various authors, e.g. \citet[][and references therein]{Shi+2015}, have argued for high accretion rates.   \citet{Ragusa+2016} instead claim that the flow of material from the inner portion of the disc through the cavity is drastically reduced for very thin and inactive disks. They argue that the higher accretion rates found in earlier studies are computational artefacts resulting from the artificially large scale heights adopted for numerical convenience.  The findings by \citet{Ragusa+2016} favor the possible long-term survival of the disk, which is needed for the scenario we propose here. 

At present, it appears unlikely that any theoretical calculation can robustly answer the question of whether a fraction of the original circumbinary material can survive as a disk until the BH merger occurs. Given this uncertainty, we simply parametrize the disk mass $M_d$ as a fraction $q_d \equiv M_d/M$ of the total binary mass.

The orbit of the binary system decays as a result of gravitational wave radiation. The timescale for this process is given by \citet{Peters1964}
\begin{equation}
t_{\rm GW} \simeq 1.1 \times10^{4}\, (a/\Rsun)^{4}M_{60}^{-3}\,\,{\rm yr}  \label{tgw}
\end{equation}
where we have assumed equal masses. Initially, the timescale for orbital decay is long compared to viscous timescale and the disk can viscously respond to the change of the orbital separation by spreading and shrinking the inner cavity.  (step 1 in \autoref{scenario}). 

Eventually, $t_{\rm GW}$ becomes shorter than the viscous timescale $t_{\rm visc} (R_{\rm in})$ and 
the evolution of the disk decouples from the inner binary 
\citep[e.g.][step 2 in \autoref{scenario}]{Milosavljevic+2005}. This happens very shortly before the merger.  
The size of the inner radius $R_{\rm in }$  of the disk can be estimated by solving $t_{\rm GW} (a) = 
t_{\rm visc} (R_{\rm in })$ where $R_{\rm in } \simeq 2a$.  The inner edge of the disk is now deep in the 
potential well of the BH binary, so must be hotter and ionized.  We therefore now use the parametrization $
\alpha_{-1} = \alpha/10^{-1}$ appropriate for an ionized disk, which gives
\begin{equation}
R_{\rm in }  \simeq
3\times 10^{10}\alpha_{-1}^{-2/5}[(H/R)_{-3}]^{-4/5}  M_{60}  \,{\rm cm}.
\label{rin}
\end{equation}

\paragraph{Response to sudden mass loss and recoil at coalescence} 

When the binary black hole coalesces,  the emission of gravitational waves produces a sudden reduction of the total central mass and it imparts an impulsive kick to the newly formed merged black hole \citep[illustrated in step 3 in \autoref{scenario}, see][]{Peres1962, Fitchett1983,  Campanelli+2007, Tichy+2008}.
\citet{Barausse+2012} and \citet[][and references therein]{Lousto+2014} provide semi--analytic expressions for the mass radiated away and the recoil velocities \citep[see also][]{Schnittman+2007,Baker+2007}.  They find significant fractions of mergers with  $v_{\rm rec} > 500 - 1000 ~{\rm km\, s^{-1}}$, depending on the orientations and magnitudes of the spins of the coalescing black holes.  These in turn bear the imprint of the spin evolution of the progenitor stars, modified by their subsequent implosion and possible birth kicks when forming black holes. These could be substantial in some cases \citep[e.g.][]{Repetto+2012}, but definitely not in all \citep{Mirabel+2003}. Post--Newtonian effects can later align the binary further in some cases \citep[e.g.][]{Schnittman2004}. For the progenitor formation scenarios for isolated binary evolution (i and ii, see the \emph{origin} section) one may expect spins with a substantial component out of the plane of the binary. The newly merged black hole resulting from such a system is generally expected to have a recoil with a substantial component directed perpendicular to the plane of the binary \citep[e.g.][]{Campanelli+2007,Lousto+2012}. 
  
Particles that remain bound to the newly formed black hole suddenly find themselves on elliptical orbits.  The gas disk is expected to be cool and thin, so the orbital motion of the gas is hypersonic and susceptible to prompt shocks. These can, in principle, produce a transient EM signature \citep{Milosavljevic+2005}. Various groups have studied some or all of these effects in the context of supermassive binary black hole mergers, where bright post--merger flares because of shocks have been suggested as possibly detectable transient EM signatures \citep[e.g.][]{Bode+2007, Lippai+2008, Schnittman+2008, Shields+2008, Megevand+2009, ONeill+2009, Krolik2010, Rossi+2010, Corrales+2010} 

Particularly large effects are expected for recoils with a substantial component $v_{\rm rec, ||}$ in the binary 
plane \citep{Rossi+2010}. If this is in the 12 o'clock direction for a counterclockwise disk, gas at the 9 o'clock position suddenly 
has a greatly reduced angular momentum with respect to the (now moving) center of mass, particularly if 
$v_{\rm rec, ||}$ is comparable to the Kepler velocity $v_K$ at the inner edge of the disk. This material falls 
deep into the potential well of the newly formed black hole and releases energy as it circularizes much 
closer to the black hole.  

\section{Electromagnetic signal} \label{sec:pred}

We give crude estimates of the main characteristics of the resulting EM emission based on simple scaling relations calibrated against the numerical simulations by  \citet{Rossi+2010}.  We caution that the numerical estimates below are subject to large uncertainties.

\noindent
\emph{1. Time delay and duration:} Heating of the disk by shocks occurs on a characteristic timescale of order the dynamical time
\begin{equation}
t_{\rm dyn} \sim {GM\over v^3} \sim 2.2{M_{60}\over v_{3}^3}~{\rm hrs},   
\label{t}
\end{equation}
\noindent where $v_{3} = v/(10^3\, {\rm km\, s^{-1}})$ and  $v$ is the greater of the Keplerian velocity at the inner orbit and the recoil velocity imparted to the center of mass, i.e. 
$
v \equiv \max \{v_K (R_{\rm in}), v_{\rm rec}\}.
$
\noindent  
The light--travel time required to tell  the disk about the GW event is much shorter than  $t_{\rm dyn}$. Therefore, $t_{\rm dyn}$ describes the delay with respect to the GW event and energy deposition in the disk. This means that we expect a minimum time delay of hours between the merger and the EM signal, if photons are radiated away promptly. The time delay becomes longer, and can be years if a significant fraction of the radiation is trapped, as we discuss in {\it 3.} below.

\noindent
\emph{2. Disk dimensions:} We define a typical radius  $R_{\rm rec}$ such that the Kepler velocity at this radius is equal to the recoil velocity  
\begin{equation}
R_{\rm rec} \equiv {GM\over v_{\rm rec}^2} \sim  10 {M_{60}\over v_{3}^2}\Rsun. 
\end{equation}
In the typical case we expect $v_{\rm rec} < v_K (R_{\rm in})$,  so that the inner radius of the disk falls  within this radius (cf. \ref{rin}).  The outer disk radius is much larger than when the disk was formed (i.e. $R_{\rm out} \gtrsim \Rsun$, as it must absorb the angular momentum of the gas which has spiraled inwards. Therefore we expect the outer radius to be significantly larger than $R_{\rm rec}$.

\noindent  
\emph{3. Peak luminosity}: the expected rate of dissipation of kinetic energy is 
\begin{eqnarray}
L &\sim& f  {M_{\rm d}v^2\over t_{\rm dyn}}  \sim f \,\,
{v^5\over G} {q_d}
\end{eqnarray}
where $f$ is a scaling factor. We calibrate the factor $f\sim 0.1$ against the numerical simulations by \citet{Rossi+2010},  which assume an angle of  $\theta = 15\degree$ between the recoil direction and the orbital plane
\begin{eqnarray}
L &\sim& 5 \times 10^{42} \left(\frac{f}{0.1}\right) \,\, v_3^5 \left({q_{\rm d} \over 10^{-3}} \right)  \,{\rm erg\,s^{-1}},  
\label{L}
\end{eqnarray} 
because the dissipation lasts several dynamical times \citep[cf. Fig 22 of ][]{Rossi+2010}.  
The peak EM luminosity depends on how promptly the radiation appears.   The luminosity (\ref{L}) is highly super--Eddington for any stellar--mass merger.  If the shocked disc reacts by expanding homologously it would retain a high optical depth until significantly expanded. This means that it can take years for the photons to appear, which would  strongly reducing the luminosity potentially rendering the signal undetectable. If instead the photons are released on a timescale that is comparable to the few--hour energy production time, then the peak luminosity of the EM signal is comparable to the rate of energy dissipation given in \ref{L}. This would give EM fluxes of up to $10^{-13}v_3^5(f/0.1)\,{\rm erg\,s^{-1}}$ at the $\sim 500$ Mpc distance of GW150914. This is potentially detectable at the expected photon energies (X--rays, see below).

The peak EM luminosity also depends very sensitively on the kick velocity ${\bf v}$ and its direction, and on the disk--to--black hole mass ratio $q_d$. Note however that it is otherwise independent of the black hole mass. Remarkably, this means that the signal of stellar--mass events can potentially be as luminous as for mergers of supermassive black holes (but note that a more massive merger lasts longer (cf. Eq.~\ref{t}), and so emits more energy in total). 

\noindent 
\emph{4. Spectral energy distribution:} The characteristic temperature $T$ associated with the burst satisfies
\begin{equation}
T_b < T < T_s
\label{T}
\end{equation}
where $T_b$ is the blackbody temperature and $T_s$ is the shock temperature given by 
\begin{eqnarray}
T_b &\equiv& \left({L\over 2\pi \sigma R_v^2}\right)^{1/4}  \label{tb} \\
&\sim& 3\times 10^6\,\,v_{3}^{9/4}\,M_{60}^{-1/2}\,  \left( {q_{\rm d} \over 10^{-3}} \right)^{1/4}\,\left( {f \over 0.1} \right)^{1/4}~{\rm K}, \nonumber \\
T_s &\equiv& {3\mu m_Hv^2\over 16 k} 
\sim 2\times 10^7\,v_3^2~{\rm K} \label{ts}.
\end{eqnarray}
~\\
Comparing (\ref{tb}, \ref{ts}) we see that for $v_3 \sim 1$ and assuming prompt emission the EM signal is likely to peak in medium--energy X--rays.  Detection may also be possible at longer wavelengths, for example as the Rayleigh--Jeans tail in the infrared. This may be important if the emission is heavily reddened.

Given the luminosity estimate (\ref{L}) and the 
timescale (\ref{t}) it seems possible that events like these may already have been observed, but 
ascribed to variability in unidentified AGN.  

\emph{5. Additional spectral features}

The EM luminosity can be greatly enhanced for kicks in the orbital plane where  $v_{\rm rec} \lesssim v_K (R_{\rm in})$ and the material falls deep into the potential of the merged black hole. This gives red-- or  blueshifts of the same order (say $\sim 1000\,{\rm km\, s^{-1}}$) for the EM emission, which can in principle be used to probe the post merger recoil. Further, the composition of the material of the disk may be enriched if the material results from ejecta that were processed by nuclear burning in the progenitor stars. For the closest events this might eventually lead to detectable spectral diagnostics.

\section{Discussion}

Our estimates are subject to various significant uncertainties. These include, but are not limited to, the evolution of the circumbinary disk and its response to heating by tides, spiral waves and cosmic rays, the pre--merger spins of the black holes, the size and direction of the recoil, and how promptly the dissipated energy appears from the shocked disk. 

Given these large unknowns, we cannot state with certainty whether the signals are observable or not. Further detailed simulations will be needed for more accurate predictions, and even these face the difficulty that the behavior of the disk viscosity over time is inherently uncertain. A definitive answer will only come from observations. 

\paragraph{Observing the signal} 
 If the EM signal appears a few hours after the GW merger, a fairly accurate position is needed for suitable instruments to scan the error box on a similar timescale. This is not possible with the currently available combination of LIGO and SWIFT alone, but will become increasingly feasible as additional GW detectors come online and allow accurate triangulation of GW events. A particularly attractive possibility is that eLISA may be able to detect black hole binaries several years {\emph before} coalescence \citep{Sesana2016}, allowing targeted campaigns. Our estimates  (\ref{L}, \ref{tb}, \ref{ts})  of the  luminosity and temperature, and the possible radial velocity signal, suggest that searches from  X--rays to infrared could be fruitful. There are a large number of current, planned, and projected instruments which would be suitable for this (e.g. ZTF, ATLAS, GOTO, BlackGem, LSST, SVOM, Einstein Probe).

\paragraph{Other possible EM signals} Recently, various  groups proposed the possibility of other types of observable EM signals, encouraged by the tentative Fermi signal.  \citet{Perna+2016} also consider the possibility that material ejected in the second weak SN explosion might form a dead disk, which we also consider, but our scenario differs from theirs in various aspects. They assume a minidisk around one of the individual black holes, heated by  tidal torques and shocks during the pre-merger phase. This rapidly consumes the disk and powers a short gamma--ray burst coinciding with the merger event \citep[see also][]{Murase+2016}.  \citet{Kimura+2016} argue to the contrary that such a minidisk would be heated and activated a few years before the merger event and lead to a precursor event instead. Our scenario differs in the disk geometry, which is circumbinary in our picture.  As a result the brightening arises from perturbations of the circumbinary disk itself, on its dynamical timescale. This is not directly related to viscous accretion on to the newly formed black hole, where the timescale is much longer. Other suggestions beyond those mentioned in the Introduction include signals from the merger of charged black holes \citep{Zhang2016, Fraschetti2016, Liebling+2016}. 

\section{Conclusions}  \label{sec:dis}

We have suggested that mergers of stellar--mass black hole binaries driven by gravitational wave emission potentially produce EM counterparts if a low--mass circumbinary disk survives until the coalescence. We argue that such a disk responds to sudden mass loss and recoil of the GW merger leading to shock, which heats it within hours of the merger.  Whether the signal is detectable is uncertain. If the optical depth through the disk is high the prompt emission consists only of photons from its optically thin outer layers, while the majority may take years to emerge. If by some mechanism, photons are released promptly the resulting EM signal will probably appear similar to background AGN events, most likely in medium--energy X--rays, possibly extending to the infrared, and last at least a few hours. They would become identifiable once triangulation by future GW detectors produces error boxes small enough that prompt--response EM instruments can scan them on this timescale. In certain cases they may be pre--detected years in advance once the Evolved Laser Interferometer Space Antenna comes online. We also expect significant radial velocities ($\sim 1000\, {\rm km\,s^{-1}}$) in certain cases, which might be detectable in a rapid optical follow-up. 

Despite the large uncertainties, we suggest that the potential rewards of a successful detection of the signal discussed here justify the effort of coordinated EM campaigns. It would place unique constraints on the binary evolution before the merger, and thus provide crucial information about 
the contribution of various progenitor channels. Detection of a radial velocity, as suggested above, would also 
constrain the recoil predicted by the GW data, and so directly test strong--field predictions of General 
Relativity.

\vspace{0.05in} 
{\bf Acknowledgements.} We thank the anonymous referees for helpful comments, and A. Buonanno, Z. Haiman, A. MacFadyen, Y. Levin, I. Mandel, C. Miller, Ch. Nixon, P. O'Brien, M. Renzo, E. Rossi and R. Wijers for stimulating discussions. SdM acknowledges support by a Marie Sklodowska-Curie Action (H2020 MSCA-IF-2014, project id 661502) and National Science Foundation under Grant No. NSF PHY11-25915. Astrophysics research at the University of Leicester is supported by an STFC Consolidated Grant.  


\label{lastpage}
\end{document}